\documentclass[seceq]{ptptex}

\usepackage{graphicx}

\newcommand{\fslash}[1]{\mbox{$\!\not\!#1$}}

\newcommand{\be}{\begin{equation}}
\newcommand{\ee}{\end{equation}}
\newcommand{\ba}{\begin{eqnarray}}
\newcommand{\ea}{\end{eqnarray}}

\title{
A Faddeev Calculation for Pentaquark $\Theta^+$ in Diquark
Picture\\with Nambu-Jona-Lasinio Type Interaction

}

\author{
  H.
 \textsc{Mineo}$^{1,2,}$\footnote{ e-mail address:
 mineo@gate.sinica.edu.tw}, J.A. \textsc{Tjon}$^{1,3,}$\footnote{e-mail
 address: J.A.Tjon@phys.uu.nl}, K.
 \textsc{Tsushima}$^{4,5,}$\footnote{ e-mail address:
 tsushima@usal.es},  and {\underline{Shin Nan   \textsc{Yang}$^{1,}$}}\footnote{ e-mail
 address: snyang@phys.ntu.edu.tw}}

\inst{
$^1$ Department of Physics, National Taiwan University, Taipei
10617, Taiwan\\
$^{2}$ Institute of Atomic and Molecular Sciences, Academia Sinica,
\\P.O.Box 23-166, Taipei 10617, Taiwan\\
$^{3}$KVI, University of Groningen, The Netherlands\\
$^4$National Center for Theoretical Sciences, National Taiwan
University,\\ Taipei 10617,
Taiwan\\
$^5$Grupo de F\'{\i}sica Nuclear and IUFFyM, Universidad de
Salamanca,\\ E-37008 Salamanca, Spain}

\abst{
A Bethe-Salpeter-Faddeev (BSF) calculation is performed for the
pentaquark $\Theta^+$ in the diquark picture of Jaffe and Wilczek in
which $\Theta^+$ is a diquark-diquark-${\bar s}$ three-body system.
 Nambu-Jona-Lasinio (NJL)
model is used to calculate the lowest order diagrams in the
two-body scatterings of ${\bar s}D$ and $D D$. With the use of
coupling constants determined from the meson sector, we find that
${\bar s}D$ interaction is attractive while $DD$ interaction is
repulsive. However, we can not find a bound $ \frac 12^+$
pentaquark state. Instead, a bound pentaquark $\Theta^+$ in $
\frac 12^-$ channel is obtained with unphysically strong vector
mesonic coupling constants.}

\begin{document}

\maketitle

\section{Introduction}\label{sec1}
Experimental situation on the existence of $\Theta^+$ is currently
filled with conflicting positive and negative evidence and the
question is still not yet fully settled \cite{Nakano06}.

Even if it turns out that $\Theta^+$ does not exist, it would still
be   theoretically interesting to understand the absence of such an
exotic. Many theoretical approaches have been used, including quark
models, QCD sum rules, and lattice QCD, in addition to the chiral
soliton model,  to understand the properties and structure of
$\Theta^+$ since its first sighting \cite{LEPS03}. One of the most
intriguing theoretical ideas suggested for $\Theta^+$ is the diquark
picture of Jaffe and Wilczek (JW) \cite{Jaffe03} in which $\Theta^+$
is considered as a three-body system consisted of two scalar,
isoscalar, color $\bf {\bar 3}$ diquarks ($D$'s) and a strange
antiquark $(\bar s)$. It is based, in part, on group theoretical
consideration. It would hence be desirable to examine such a scheme
from a more dynamical perspective.

It is known that diquark arises naturally from  Nambu-Jona-Lasinio
(NJL) model, an effective quark theory in the low energy region
\cite{NJL}. NJL model conveniently incorporates chiral symmetry and
its spontaneously breaking which dictates the hadronic physics at
low energy. Models based on NJL type of Lagrangians have been very
successful in describing the low energy meson physics. Based on
relativistic Faddeev equation, the NJL model has also been applied
to the baryon systems \cite{Huang94,Mineo99}. It has been shown
that, using the quark-diquark approximation, one can explain the
nucleon static properties  and the qualitative features of the
empirical valence quark distribution reasonably well \cite{Mineo99}.
Consequently, we will employ NJL model to describe the dynamics of a
$(\bar sDD)$ three-particle system in Faddeev formalism. We use
relativistic equations to describe both the three-particle and its
two-particle subsystems, namely, the Bethe-Salpeter-Faddeev (BSF)
equation \cite{Rupp88} and Bethe-Salpeter (BS) equations. In
practice, Blankenbecler-Sugar reduction scheme is used to reduce the
four-dimensional integral equation into three-dimensional ones.
\section{$\bf SU(3)_f$ NJL model and the diquark}\label{sec2}
The SU(3)$_f$ NJL model is a chirally symmetric four-fermi contact
interaction Lagrangian. With the use of Fierz transformations, the
original NJL interaction Lagrangian $\cal L_I$ can be rewritten, for
the $q{\bar q}$ channel, as

\begin{eqnarray} {\cal L}_{I,q{\bar q}} &=& G_1\left[ ({\bar \psi}\lambda^a_f
\psi)^2 -({\bar \psi}\gamma^5 \lambda^a_f \psi)^2 \right] -G_2
\left[ ({\bar \psi} \gamma^{\mu}\lambda^a_f\psi)^2 +({\bar
\psi}\gamma^{\mu} \gamma^5\lambda^a_f \psi)^2
\right]\nonumber\\
&-&G_{3} \left[ ({\bar \psi}\gamma^{\mu}\lambda^0_f  \psi)^2 +({\bar
\psi}\gamma^{\mu}\gamma^5 \lambda^0_f \psi)^2 \right] -G_{4} \left[
({\bar \psi}\gamma^{\mu}\lambda^0_f \psi)^2 -({\bar
\psi}\gamma^{\mu}\gamma^5 \lambda^0_f \psi)^2
\right]\nonumber\\
&+&\cdots , \label{NJLLagrangian} \end{eqnarray} where $a=0\sim 8$,
and $\lambda_f^0=\sqrt{\frac23}I$. For later use, we define
$G_5=G_2+\frac23G_v,$ with $G_v\equiv G_3+G_4.$ For the scalar,
isoscalar diquark channel,  interaction Lagrangian is given by
\begin{eqnarray}  {\cal L}_{I,s} = G_s \left[ {\bar \psi}(\gamma^5 C ) \lambda_f^2
\beta^A_c {\bar \psi}^{T}\right] \left[ \psi^T (C^{-1} \gamma^5
)\lambda_f^2
 \beta^{A}_c\psi \right],
\label{L_Is} \end{eqnarray}  where
$\beta^A_c=\sqrt{\frac32}\lambda^A (A=2,5,7)$ corresponds to one of
the color ${\bar 3}_c$ states. $C=i\gamma^0\gamma^2$ is the charge
conjugation operator, and $\lambda's$ are the Gell-Mann matrices.

The constituent quark and diquark masses can be obtained from the
gap equation and t-matrix of the diquark. Since we are only
interested in a qualitatively study of the interactions inside
$\Theta^+$, we will use the empirical values of the constituent
quark masses $M_{u,d}=400$ MeV, $M_s=600$ MeV, and the diquark mass
$M_D=600$ MeV as obtained in Ref. \cite{Mineo}.

\section{Two-body interactions for  $\bf\bar
sD$ and $DD$ channels} \label{sec3}
In the JW model for $\Theta^+$
\cite{Jaffe03}, symmetry consideration requires that the the spatial
wave function of the  two scalar-isoscalar, color $\bf \bar 3$
diquarks must be antisymmetric and the lowest possible state is
$p$-state. Since $\Theta^+$ is of $J^P={\frac 12}^+$, $\bar s$ would
be in relative $s$-wave to the $DD$ pair. Accordingly, we will
consider only the configuration where $\bar sD$ and $DD$ are in
relative $s$- and $p$-waves, respectively.

Fig. 1 shows the lowest order diagram, i.e., first order in ${\cal
L}_{I,q{\bar q}}$ in ${\bar s}D$ scattering. Trace properties in
Dirac and flavor space limit the vertex $\Gamma$ to only the
vector-isoscalar term, $({\bar \psi}
\gamma^{\mu}\lambda^0_f\psi)^2$. For the $DD$ interaction, the quark
rearrangement diagram gives no contribution because of its color
structure. The lowest order non-vanishing diagram of the first order
in ${\cal L}_{I, q\bar q}$ is given in Fig. 2 where only the direct
term is shown. The corresponding exchange diagram vanishes again
because of the color structure.

\begin{figure}[htbp]
   \begin{minipage}{.4\textwidth}
    \hfill
    \includegraphics[width=5.0cm]{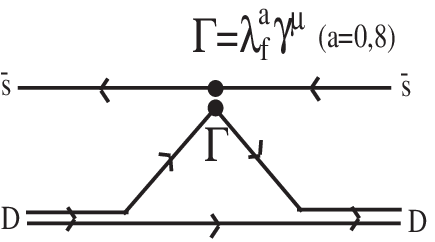}
    \caption{${\bar s}$D potential
of the lowest order in ${\cal L}_{I,q{\bar q}}$. }
   \end{minipage}
   \begin{minipage}{.15\textwidth}
    \hfill
   \end{minipage}
   \begin{minipage}{.4\textwidth}
     \includegraphics[width=5.0cm]{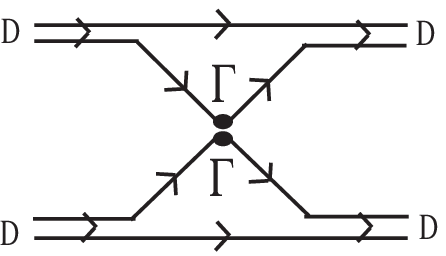}
    \caption{Lowest order diagrams in $DD$ scattering.}
  \end{minipage}
\end{figure}
With the use of the interaction Lagrangians of Eqs.
(\ref{NJLLagrangian}-\ref{L_Is}), we obtain the following driving
terms of Fig. 1 and 2, in the BS equations for $\bar sD$ and $DD$
two-particle systems, \ba <{\bar s}_fD_f| V|{\bar s}_iD_i>
&=&(-{\bar v}(p_{{\bar s}i}))(-i V_{{\bar s}D})
(p_{Di},p_{Df})v(p_{{\bar s}f}),\nonumber
 \label{VsbarD} \ea with \be {V}_{{\bar
s}D}=\frac{64}{3} G_v F_v(q^2) \tilde{V}_{{\bar s}D}(p_{Di},p_{Df}),
\hspace{1.0cm} \tilde{V}_{{\bar s}D}({p}_{Di},{p}_{Df})
=(\fslash{p}_{Di}+\fslash{p}_{Df})/2 \label{tildeVsbD}\ee and, 
\ba
-iV_{DD}(\vec{p}_{Di},\vec{p}_{Df})=128i \left [G_1 F_s^2(q^2) -G_5
(p_{D1i}+p_{D1f})\cdot(p_{D2i}+p_{D2f})F_v^2(q^2)\right ],
 \label{128i}
\ea 
where $p_{\bar s}$ and $p_{D}$ denote the four-momentum of the
$\bar s$-quark and diquark etc. The $F_v $ and $F_s $ are the vector
and scalar form factors of the scalar diquark. For simplicity, we
will assume  that both take the dipole form, $
(1-q^2/\Lambda^2)^{-2}$, with $\Lambda=0.84$ GeV. In the NJL model
calculation with the Pauli-Villars cutoff \cite{Mineo}, the coupling
constants are related to the mesonic coupling constants by
$G_1=G_{\pi}/2$, $G_2=G_{\rho}/2$ and $G_5=G_{\omega}/2$ which give
$G_v=-0.78$ GeV$^{-2}$.  
We remark that the sign of $G_v$ is
definitely negative since omega meson is heavier than the rho meson.

The potential matrix elements of Eqs. (\ref{VsbarD}-\ref{128i}) can
then be used in the scattering equations obtained with the use of
Blankenbecler-Sugar three-dimensional reduction scheme \cite{Rupp88}
for the BS equation for both the $\bar sD$ and $DD$ systems. The
resultant  scattering equations are solved to obtain the two-body
$t-$matrix elements and the phase shifts.

\section{Results for $\Theta^+$ and discussion}\label{sec4}
Our results for the phase shifts are shown in Fig. 3(a). We see that
the s-wave phase shifts for $\bar sD$ is positive which indicates
the interaction is attractive, while the
 p-wave $DD$ interaction is
repulsive since their phase shift is negative. In Fig. 3(b) we show
the $G_v$ dependence of the two-body $\bar sD$ binding energy. We
see that with the type of interaction   constructed in
Sec.~\ref{sec3}, a $\bar sD$ bound state begins to appear only when
$G_v$ becomes less than $-5\sim -6$ Gev$^{-2}$, far too negative as
compared to the physical value of -0.78 Gev$^{-2}$ determined from
the $\rho-\omega$ mass difference.

\begin{figure}[tbh]
\centering
\includegraphics[width=5.5cm]{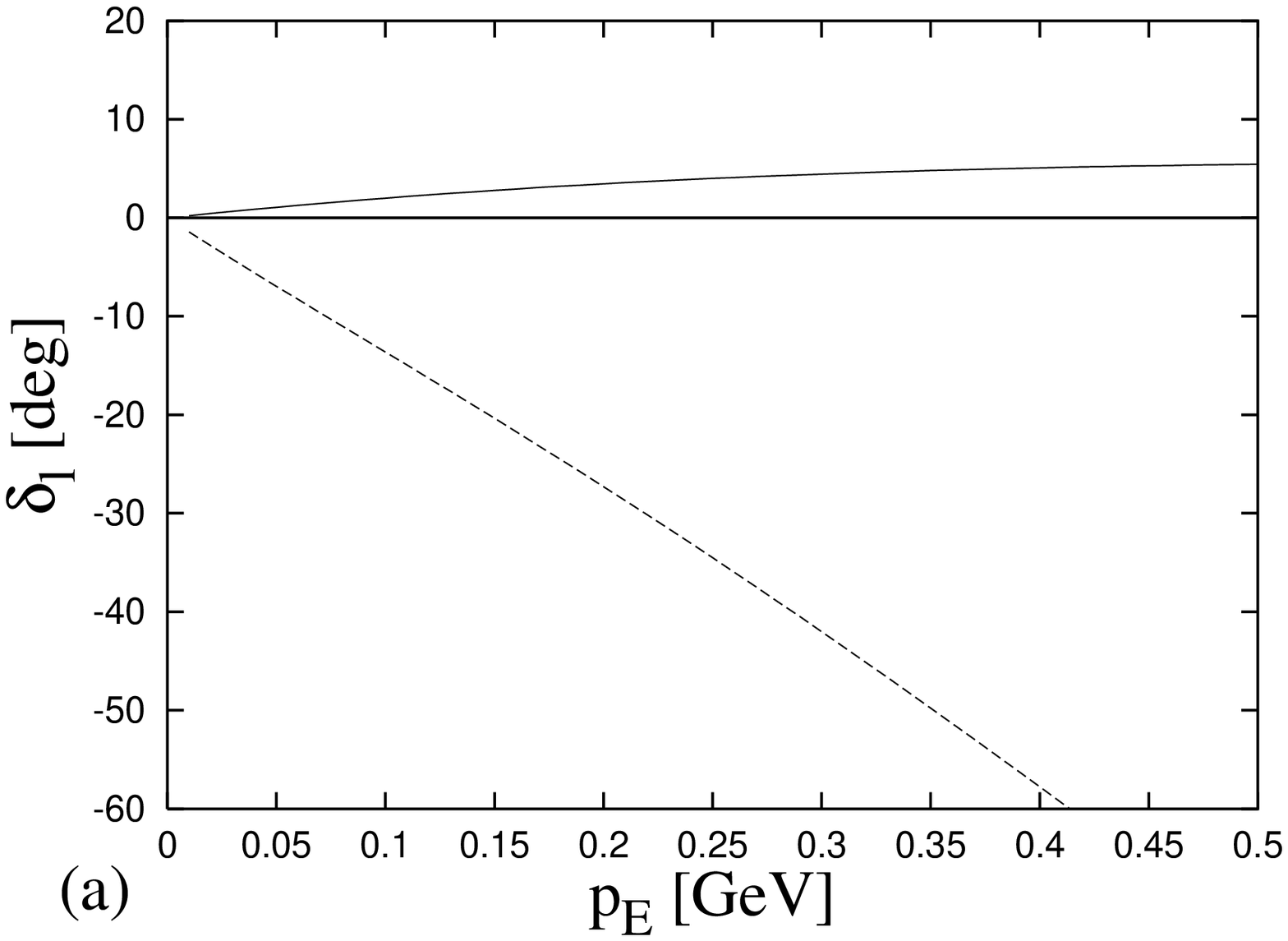}
\hspace{1.0cm}
\includegraphics[width=5.5cm]{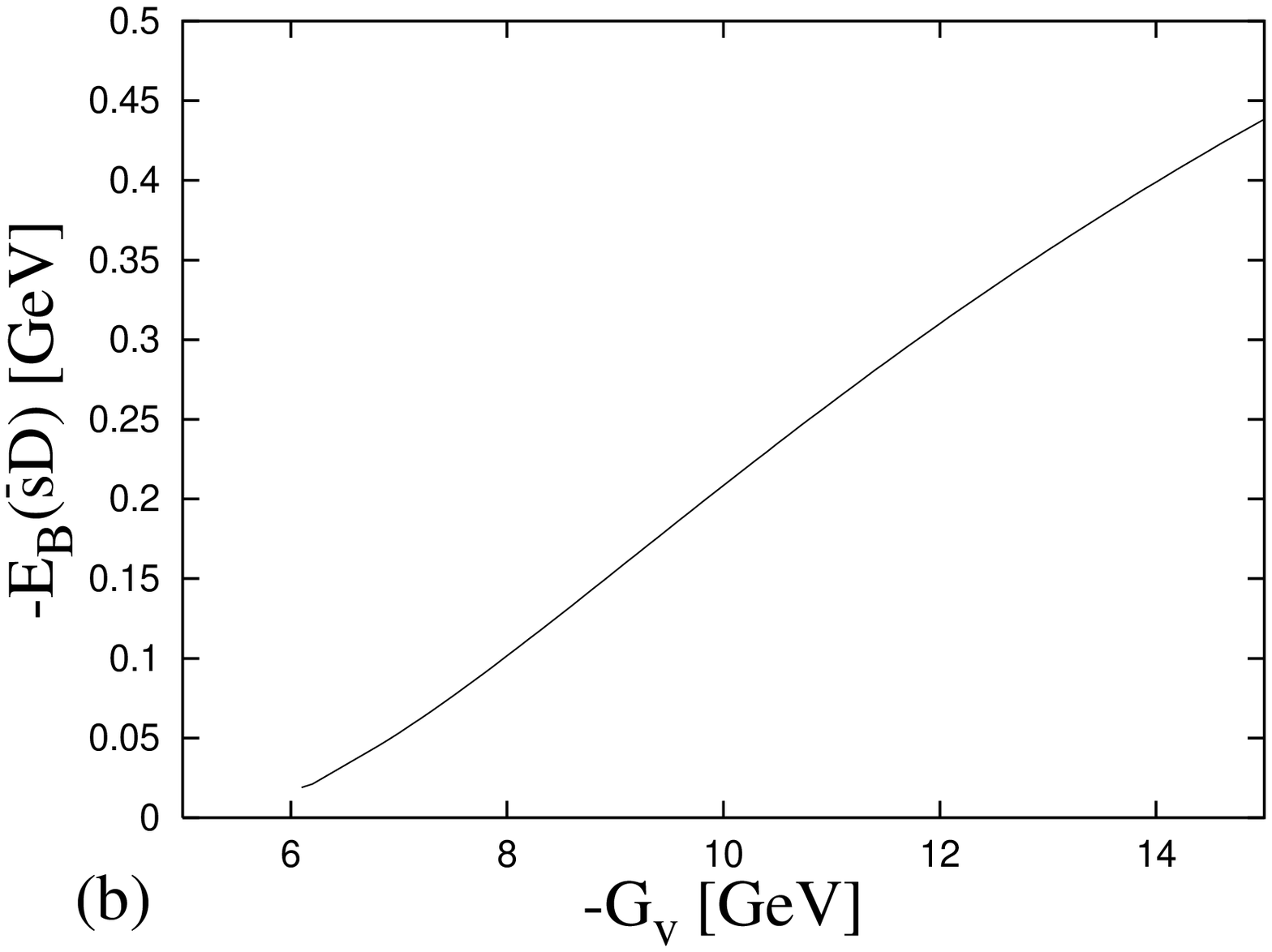}
\caption{(a) Phase shifts $\delta_l$ for the ${\bar s}D$ in s-wave
with $G_v=-0.78$ GeV$^{-2}$ (solid line) and DD in p-wave (dashed
line). (b) $G_v$ dependence of the ${\bar s}$D binding energy.}
\end{figure}

The three-body BSF equation \cite{Rupp88} takes the same form as the
nonrelativistic one, \be T_i(s)=t_i(s)+t_i(s)G_0(s)\left
[T_j(s)+T_k(s)\right ],\label{Faddeev}\ee where $G_0$ is the free
three-particle Green's function and $t_i(s)$ is the two-particle
$t-$matrix of particles $j$ and $k$ with $(i,j,k)$ being a cyclic
permutation of $(1,2,3)$. If one uses the Blankenbecler-Sugar
approximation for $G_0$ and the two-body $t-$matrix elements
obtained in Sec. \ref{sec3}, then the homogeneous equation of Eq.
(\ref{Faddeev}) can be solved \cite{Amazadeh66} to look for a
possible three-body $\bar sDD$ bound state. We could not find a
bound pentaquark in $J^P=\frac 12^+$ channel. However, we do see a
pentaquark in $J^P=\frac 12^-$ channel when $G_v$ becomes less that
$\sim -8.0$ GeV$^{-2}$. Pentaquark binding energy $E_B(5q)$ grows
from 77 to 505 MeV as $G_v$ decreases from -8.0 to -14.0 GeV$^{-2}$.
The effect of consierdable weaker attraction in the $J^P=1/2^+$
channel is caused by the spectator particle being in a p-wave state.

\section*{Acknowledgements}
One of the authors (S.N.Y.) thanks the Yukawa Institute for
Theoretical Physics at Kyoto University, for warm hospitality
extended to him during the YKIS2006 on "New Frontiers on QCD".

\end{document}